\documentclass[fleqn,twoside]{article}
\usepackage{espcrc2}

\usepackage{graphicx}
\usepackage[figuresright]{rotating}

\newcommand{\AmS}{{\protect\the\textfont2
  A\kern-.1667em\lower.5ex\hbox{M}\kern-.125emS}}

\title{$A_4$ Group and Tri-bimaximal Neutrino Mixing -- A Renormalizable Model}

\author{Xiao-Gang He\address[MCSD]{Department of Physics and Center for Theoretical Sciences\\
        National Taiwan University, Taipei, Taiwan}%
        }

\begin{document}

\begin{abstract}
The tetrahedron $A_4$ group has been widely used in studying
neutrino mixing matrix. It provides a natural framework of model
building for the tri-bimaximal mixing matrix. In this class of
models, it is necessary to have two Higgs fields, $\chi$ and
$\chi'$, transforming under $A_4$ as 3 with one of them having
vacuum expectation values for the three components to be equal and
another having only one of the components to be non-zero. These
specific vev structures require separating $\chi$ and $\chi'$ from
communicating with each other. The clash of the different vev
structures for $\chi$ and $\chi'$ is the so called sequestering problem. In this work, I
show that it is possible to construct renormalizable
supersymmetric models producing the tri-bimaximal neutrino mixing
with no sequestering problem. \vspace{1pc}
\end{abstract}

\maketitle

The current data from neutrino oscillation experiments\cite{data}
can be described by three neutrino mixing. The mixing matrix $V$
can be well fitted by the tri-bimaximal mixing of the
form\cite{hps}
\begin{equation}
V_{tri-bi}=\left(
\begin{array}{rrr}
{\frac{2}{\sqrt{6}}} & {\frac{1}{\sqrt{3}}} & 0 \\
-{\frac{1}{\sqrt{6}}} & {\frac{1}{\sqrt{3}}} & -{\frac{1}{\sqrt{2}}} \\
-{\frac{1}{\sqrt{6}}} & {\frac{1}{\sqrt{3}}} &
{\frac{1}{\sqrt{2}}}
\end{array}
\right)P.  \label{tri-bi}
\end{equation}
Here $P
= Diag(e^{i\alpha_1}, e^{i\alpha_2}, e^{i\alpha_3})$ is a Majorana phase matrix.
Since an overall phase does not play a role in any physical
process, only two of the $\alpha_{1,2,3}$ are physically
independent.

The tri-bimaximal form for neutrino mixing was first proposed by
Harrison, Perkins and Scott\cite{hps}, and further studied by Xing\cite{hps}.
Also independently proposed by He and Zee\cite{hps}.
Many theoretical efforts have been made to produce such a mixing
pattern. Among them theories based on $A_4$ symmetry provide some
interesting examples\cite{A4,babu,ray}. Most of the attempts made
in the literature assumed certain vacuum expectation value (vev)
structures for Higgs fields without specific renormalizable models
to realize them. Attempts to build renormalizable models have been
made in Refs.\cite{babu,ray}. Here I construct a realistic renormalizable model
with supersymmetry (SUSY) which produces the tri-bimaximal mixing.

In addition to the standard $SU(3)_C\times SU(2)_L\times U(1)_Y$
gauge symmetry, this model has additional global symmetries
$A_4\times Z_4\times Z_3\times Z_2$ acting on various fields.
Under the global symmetries in order of
$A_4$, $Z_4$, $Z_3$, and $Z_2$, the relevant lepton and Higgs fields transform as:
$ L(3,1,0,1)$, $e^c(1+1'+1'',0,2,1)$, $
N^c(3,0,1,1)$, $E(3,0,1,1),\;E^c(3,2,2,1)$, $
H_u(1,3,2,0)$, $H_d(1,1,$ $1,0),\;\chi(3,0,0,0),\;\chi'(3,0,1,0),
\;S(1,0,0,0),\;S'(1,$ $0,1,0),\;S''(1,2,0,0)$.

The $A_4$ group is the tetrahedron group. It has 12 elements with
4 inequivalent representations 1, $1'$, $1''$ and 3. The
multiplication rules of these representations are $1\times 1 = 1$,
$1\times 1'=1'$, $1\times 1''=1''$, $1'\times 1''=1$, $1'\times 1'
= 1''$, $1''\times 1'' = 1'$, $3\times 3 = 1 +1'+1''+3+3$. The
$1$, $1'$, $1''$ and the two 3's formed from two 3's $a=(a_1,a_2,a_3)$
and $b=(b_1,b_2,b_3)$ are given by
\begin{eqnarray}
&&1: a_1b_1 +a_2b_2 + a_3b_3,\nonumber\\
&&1': a_1b_1 +\omega a_2b_2 + \omega^2 a_3b_3,\nonumber\\
&&1'': a_1b_1 + \omega^2 a_2b_2 + \omega
a_3b_3,\nonumber\\
&&3_s: (a_2b_3+a_3b_2,a_3b_1+a_1b_3,a_1b_2+a_2b_1),\nonumber\\
&&3_a: (a_2b_3-a_3b_2,a_3b_1-a_1b_3,a_1b_2-a_2b_1).\nonumber
\end{eqnarray}
The $Z_n$ charge $N$ in the above are defined as $exp[i2N\pi/n]$.

The superpotential relevant to lepton masses are given by:
$W_Y=M_E E_i E^c_iS'' + f_e L_i E^c_i H_d + h^e_{j}E_i
e^c_j \chi_k + {1\over
2} f_{S'} N^c_i N^c_i S' + {1\over 2}f_{ijk} N^c_i N^c_j \chi'_k +
f_\nu L_i N^c_i H_u$.
The $Z_2$ is needed to prevent $N^c \chi'^2$ term which induces
mixing between $N^c$ and $\chi'$ with a non-zero vev for $\chi'$
and causes problem in obtaining the tri-bimaximal mixing.

The vev's of $\langle H_u \rangle = v_u$, $\langle H_d \rangle = v_d$,
$\langle S'\rangle = v_{s'}$, $\langle S''\rangle =v_{s''}$,
and $\langle \chi (\chi')\rangle$ break the
gauge symmetry and also the global symmetries. If $\langle \chi_i\rangle =x_i$
and $\langle \chi'_i\rangle =x'_i$ have the following form,
\begin{eqnarray}
x_1=x_2=x_3 = v_{\chi},\;\; x'_1=x'_3=0,\;\;x'_2
=v_{\chi'},\label{struc}
\end{eqnarray}
the mass matrices $M_{eE}$ and $M_{\nu N}$ in the Lagrangian $L =
-(e, E)M_{eE}(e^c, E^c)^T-(\nu^c, N)M_{\nu N}(\nu, N^c)^T$ are
given by
\begin{eqnarray}
&&M_{eE} = \left ( \begin{array}{ll} 0&M_{eE^c}\\
M_{e^cE}&M_{EE^c}\end{array} \right ),\nonumber\\
&&M_{\nu N} =\left (\begin{array}{ll} 0&M_{\nu N^c}\\M_{\nu^c
N}&M_{NN^c}\end{array}\right ),
\end{eqnarray}
with
\begin{eqnarray}
&&M_{eE^c}= \left (\begin{array}{lll}
f_e v_d&0&0\\
0&f_e v_d&0\\
0&0&f_e v_d\end{array}\right ),\nonumber\\
&&M_{e^cE}= \left ( \begin{array}{lll}h^e_1 v_\chi &h^e_2v_\chi& h^e_3 v_\chi\\
 h^e_1 v_\chi
&h^e_2\omega v_\chi& h^e_3 \omega^2 v_\chi\\
 h^e_1 v_\chi &h^e_2\omega^2v_\chi& h^e_3 \omega
 v_\chi\end{array}\right ),\nonumber\\
 &&M_{EE^c} = \left (\begin{array}{lll}
f_Ev_{s''}&0&0\\
0&f_Ev_{s''}&0\\
0&0&f_Ev_{s''}
\end{array}
\right )\;,\nonumber\\
 &&M_{\nu N^c}=M_{\nu^c N} = \left ( \begin{array}{lll}
f_\nu v_u&0&0\\
0&f_\nu v_u&0\\
0&0&f_\nu v_u
\end{array}\right ),\nonumber\\
 &&M_{NN} = \left (
\begin{array}{lll}
f_{s'}v_{s'}&0&f_{\chi'}v_{\chi'}\\
0&f_{s'} v_{s'}&0\\
f_{\chi'}v_{\chi'}&0&f_{s'}v_{s'}
\end{array}
\right ).
\end{eqnarray}
The above results in the following form for the light lepton mass
matrices,
\begin{eqnarray}
&&M_{e} = U_L \left ( \begin{array}{lll}
m_e&0&0\\
0&m_\mu&0\\
0&0&m_\tau
\end{array}
\right )\;, \nonumber\\
&&M^{light}_\nu = m_0 \left (\begin{array}{ccc}
1&0&x\\
0&1-x^2& 0\\
x&0&1
\end{array}
\right )\;,\nonumber\\
&&U_L = {1\over
\sqrt{3}}\left ( \begin{array}{lll} 1&1&1\\
1&\omega&\omega^2\\
1&\omega^2&\omega
\end{array}
\right )\;, \end{eqnarray} where the charged lepton masses
$m_{e,\mu,\tau}$ are give by
\begin{eqnarray}
m_i=\sqrt{3}\left (f_e v_d v_\chi/f_E v_{s''}) h^e_i(1+(h^e_i
v_\chi)^2)\over (f_E v_{s''})^2\right )^{-1/2},\nonumber
\end{eqnarray}
and $ m_0 = f^2_\nu v_u^2 f_{s'}v_{s'}/(f^2_{s'}v_{s'}^2 -
f^2_{\chi'}v^2_{\chi'})$, $ x=- f_{\chi'}v_{\chi'}/f_{s'} v_{s'} =
|x|e^{i\psi}$.

Diagonalizing the lepton mass matrices, we obtain the neutrino
mixing matrix given by eq.(\ref{tri-bi}). The Majorana phase
matrix $P$ is given by:
$P=$$Diag(e^{-i\phi_1/2},~e^{-i(\phi_1+\phi_2)/2},
~e^{-i(\phi_2+\pi)/2})$ with $\phi_1 =
arg(1+x),~\phi_2=arg(1-x)$. The eigen-masses are given by $m_1 =
|m_0| | 1+x|$, $m_2 = |m_0| |1-x^2|$ and $m_3 = |m_0||1-x|$.
Both normal and inverted neutrino mass hierarchies
are allowed\cite{babu}.

In order to obtain the tri-bimaximal mixing it is crucial to
have the $\chi$ and $\chi'$ representation
to have the specific vev structure in eq.(\ref{struc}). One needs
to make sure that this vev structure is obtainable in a given
model. In the following we demonstrate that the model proposed
here can have the desired vev structure.

Non-zero vev's of the Higgs break $A_4$, but left some residual symmetries.
The vev of $\chi$ with equal value for all three components breaks
$A_4$ down to a $Z_3$ generated by $\{I,c,a\}$, and the vev of
$\chi'$ with $x'_2$ non-zero breaks $A_4$ down to a $Z_2$ generated by $\{I,r_2\}$.
Here $a,\;c,\;r_2$ are $A_4$ group elements defined in
Ref.\cite{A4}. Acting on $3$, these group elements are
represented by
\begin{eqnarray}
&&a=\left (\begin{array}{lll}0&1&0\\0&0&1\\1&0&0\end{array}\right
),\;\;c= \left (\begin{array}{lll}0&0&1\\1&0&0\\0&1&0\end{array}
\right ),\nonumber\\
&&r_2= \left
(\begin{array}{lll}-1&0&0\\0&1&0\\0&0&-1\end{array}\right ).
\end{eqnarray}

If in the Higgs potential there are terms directly involve both
$\chi$ and $\chi'$, it is not possible to have the desired vev
structure. This is the so called ``sequestering" problem. To
separate $\chi$ from communicating with $\chi'$ requires additional
constraints. This is one of the crucial roles played by SUSY in
this model. Without SUSY there is no way to forbid terms of the
form $\chi^\dagger \chi\chi'^{\dagger} \chi'$ and therefore
destroys the desired vev structure in four dimensional
renormalizable theories. With SUSY, potentials are derived from
F-terms in the superpotential and D-terms involving gauge
interactions. Terms of the type $\chi^\dagger
\chi\chi'^{\dagger} \chi'$ are forbidden
if one only allows renormalizable terms in the model.

In the model discussed here the relevant terms in the superpotential
consistent with the global symmetry imposed is given by
\begin{eqnarray}
W_V &=& \lambda_{\chi s} \chi^2 S + \lambda_\chi \chi^3 +
\lambda_{\chi' s'}\chi'^{2} S' \nonumber\\
&+&\lambda_{\chi'} \chi'^{3} + \lambda_{s} S^3 + \mu_s S^2 +
\delta_s S + \lambda_{s'}
S'^3\nonumber\\
&+&\mu_{s''}S''^2 + \lambda_{ss''} S''^2 S + \mu_{ud} H_u H_d.
\end{eqnarray}

As is well known that soft SUSY breaking terms are need to
construct phenomenologically consistent model, one needs to have
these terms here too. Adding all terms which softly break SUSY but keep
$A_4\times Z_4\times Z_3\times Z_2$ symmetries, we have
\begin{eqnarray}
V_{soft} &=& b_1 \chi^\dagger \chi + b_2 \chi'^\dagger \chi' + b_3
S^\dagger S + b_4 S'^{\dagger} S'\nonumber\\
&+&b_5 S''^\dagger S''+
 \mu_u^2 H^\dagger_u H_u + \mu^2_d H^\dagger_d H_d \nonumber\\
&+&c_1 \chi^2 S + c_2 \chi^3 + c_3\chi'^{2} S' +c_4
\chi'^{3}\nonumber\\& +& c_5 S^3 + c_6 S^2 + c_7 S + c_8
S'^3\nonumber\\
&+&c_9 S''^2 + c_{10} S''^2 S + c_{11} H_u H_d .
\end{eqnarray}
This model differs the model in Ref.\cite{babu} in that the global
symmetries are not broken by soft SUSY breaking terms.

Using the stationary conditions of the Higgs potential, we obtain
\begin{eqnarray}
&&x_2{\partial V\over \partial x_1} - x_1 {\partial V\over
\partial x_2}= -2 (x_1^2-x_2^2)(\lambda_\chi^2 x_1x_2\nonumber\\
&&+ 6
\lambda_\chi \lambda_{\chi s} v_s x_3 + c_2 x_3) =0,\nonumber\\
&&x_2{\partial V\over \partial x_3} - x_3 {\partial V\over
\partial x_2} = -2 (x_3^2-x_2^2)(\lambda_\chi^2 x_3x_2 \nonumber\\
&&+ 6
\lambda_\chi \lambda_{\chi s} v_s x_1 + c_2 x_1)=0,\nonumber\\
&&x_2{\partial V\over \partial x_1} - x_1 {\partial V\over
\partial x_2} = -2 (x_1^2-x_2^2)(\lambda_\chi^2 x_1x_3 \nonumber\\
&&+ 6
\lambda_\chi \lambda_{\chi s} v_s x_2 + c_2 x_2)=0,\nonumber\\
&&x'_2{\partial V\over \partial x'_1} - x'_1 {\partial V\over
\partial x'_2} = -2 (x'^2_1-x'^2_2)(\lambda_{\chi'}^2 x'_1x'_2
\nonumber\\
&&+ 6
\lambda_{\chi'} \lambda_{\chi' s} v_{s'} x'_3 + c_4 x'_3)=0,\nonumber\\
&&x'_2{\partial V\over \partial x'_3} - x'_3 {\partial V\over
\partial x'_2} = -2 (x'^2_3-x'^2_2)(\lambda_{\chi'}^2 x'_3x'_2 \nonumber\\
&&+ 6
\lambda_{\chi'} \lambda_{\chi' s} v_{s'} x'_1 + c_4 x'_1)=0,\nonumber\\
&&x'_3{\partial V\over \partial x'_1} - x'_1 {\partial V\over
\partial x'_3} = -2 (x'^2_1-x'^2_3)(\lambda_{\chi'}^2 x'_1x'_3\nonumber\\
&& + 6
\lambda_{\chi'} \lambda_{\chi' s} v_{s'} x'_2 + c_4 x'_2)=0.
\end{eqnarray}
From the above one sees clearly that it is possible to have $\langle \chi \rangle$
to be of the form $x_1 = x_2=x_3 = v_{\chi}$, and $\langle \chi'\rangle$ to be
of the form $x'_1 = x'_3 = 0$, $x'_2 = v_{\chi'} $, at the minimal of the potential. The correct vev
structure to produce the tri-bimaximal neutrino mixing pattern has, therefore, been obtained.

I finally comment on the quark mixing.
If the quark fields are assigned under the
$A_4\times Z_4\times A_3\times Z_2$ as: $Q(1,1,0,1)$, $U^c(1,0,1,1),\;D^c(1,2,2,1)$,
one would obtain a superpotential,
$W_Q = Q\lambda_U H_u U^c + Q\lambda_D H_d D$. This superpotential then gives an unconstrained quark mixing.
Efforts have been made in Ref.\cite{ray} to explain the small off-diagonal elements in
quark mixing
by requiring that they are zero at tree level, but generated at loop levels.
The model constructed here has a simpler Higgs sector, although less predictions for quark mixing.

I thank Babu, Keum, Volkas and Zee for collaborations on related subjects reported here.
This work is partly supported by NSC and NCTS.

.
\end{document}